\documentclass[aps,prl,reprint,superscriptaddress,longbibliography,nofootinbib]{revtex4-1}
\usepackage{lipsum}

\usepackage{amsfonts}
\usepackage{amsmath}
\usepackage{tikz}
\usetikzlibrary{arrows,calc}
\usetikzlibrary{patterns,snakes}

\newcommand{\bea}{\begin{eqnarray}}
\newcommand{\eea}{\end{eqnarray}}

\begin{document}

%\preprint{}

\title{Toda chain flow in Krylov space}


\author{Anatoly Dymarsky}
\affiliation{Department of Physics and Astronomy, \\ University of Kentucky, Lexington, KY 40506\\[2pt]}
\affiliation{Skolkovo Institute of Science and Technology, \\ Skolkovo Innovation Center, Moscow, Russia, 143026\\[2pt]}
\author{Alexander Gorsky}
\affiliation{Institute for Information Transmission Problems, \\ Moscow, Russia, 127051\\[2pt]}
\affiliation{Moscow Institute for Physics and Technology, \\ Dolgoprudnyi, Russia, 141700 \\[2pt]}


\date{\today}
\begin{abstract}
We show in full generality that time-correlation function of a physical observable analytically continued to imaginary time is a tau-function of integrable Toda hierarchy. Using this relation we show that the singularity along the imaginary axis, which is a generic behavior for quantum non-integrable many-body system, is due to delocalization in Krylov space.
%which reflects chaos in an underlying many-body system, is due to delocalization of the operator in Krylov space. 
\end{abstract}

\maketitle

Time-correlation function of local operators is one of the standard probes of quantum many-body physics. It characterizes system's linear response and transport. With an exception of a few integrable models,  the explicit form  of the time-correlation function is not known, and a variety of methods have been devised  to describe its behavior in different limits. Among them is the recursion method \cite{doi:10.1080/00018736700101495,mori1965continued,haydock1980recursive}, which is commonly used for  analytic and numerical approximations. In this Letter we show that the recursion method should be understood as a part of a more general construction, defining Toda chain flow in Krylov space. In particular,
in full generality, time-correlation function of a physical observable, analytically continued to Euclidean (imaginary) time, is a tau-function of Toda hierarchy. Previously known examples \cite{izergin1984quantum,Bettelheim_2006,gmmm,Grassi:2019txd}   when a quantum correlation function was related to a classical tau-function of an integrable  hierarchy were in the context of very particular  integrable or supersymmetric theories. Here we consider generic dynamical systems and  generic observables.

One of the open questions of quantum many-body dynamics is to characterize chaotic behavior. This question connects very different  pursuits from quantum gravity \cite{shenker2014black} to mesoscopic  thermodynamics  \cite{doi:10.1080/00018732.2016.1198134}. Recently it was suggested \cite{Parker_2019} that the time-correlation function reflects underlying quantum chaotic behavior through the growth of  Lanczos coefficients defined via continued fraction expansion \cite{mori1965continued}. We apply the relation to Toda chain to elucidate this picture of chaos. We show that the singularity of time-correlation function in imaginary time, which is a generic behavior  for quantum non-integrable system \cite{alex2019euclidean}, is due to delocalization of the operator in Krylov space.

We begin by reminding the reader basics of the recursion method \cite{PhysRevB.27.7342,viswanath2008recursion}. The starting point is the time-correlation function of some operator $A$,
\bea
\label{ac}
C(t)=\langle A(t),A\rangle,
\eea
defined with the help of a Hermitian form  in the space of operators 
\bea
\label{scalarproduct}
\langle A, B\rangle\equiv {\rm Tr}(A^\dagger \rho_1 B \rho_2\rangle=\langle B,A\rangle^*.
\eea
Here $\rho_1,\rho_2$  are some  Hermitian positive semi-definite operators which commute with the Hamiltonian $H$. Therefore the adjoint action $[H,\, \, \, ]$ is self-adjoint with respect to $\langle\, \, ,\,\rangle$.  Colloquially \eqref{scalarproduct}  is a scalar product in the space of operators, with the caveat that it might be positive semi-definite rather than definite. 
For any initial $A_0=A$ we define a basis in the Krylov space via the iterative relation 
\bea
\label{iteration}
A_{n+1}=[H,A_n]-a_n A_n-b_{n-1}^2 A_{n-1},
\eea
and require $A_k$ to be mutually orthogonal. This fixes Lanczos coefficients $a_n,b_n$ to be
\bea
\label{ab}
a_n={\langle [H,A_n],A_n\rangle \over \langle A_n,A_n\rangle},\quad b_n^2={\langle A_{n+1},A_{n+1}\rangle \over \langle A_n,A_n\rangle},\,\, n\geq 0.\, 
\eea
For any Hermitian operator,  its norm defined with help of \eqref{scalarproduct} is manifestly real and non-negative. It is therefore convenient to introduce ${q_n}= \ln\langle A_n,A_n\rangle$ such that
\bea
\label{G0}
G_{nm}=\langle A_n,A_m\rangle =\delta_{nm} e^{q_n}.
\eea
In \eqref{iteration} we formally require $b_{-1}=0$.

In what follows we focus on the Euclidean time evolution,
\bea
O(t)\equiv e^{tH} O e^{-tH},  
\eea 
where  $t$ is  Euclidean (imaginary) time. An operator evolved in conventional (Minkowski) time  is  $O(-it)$.
With help of \eqref{iteration} adjoin action of $H$ in the Krylov basis $A_n$ can be represented by a Jacobi matrix $L$,
\bea
[H,A_n]&=&\sum_m L_{nm}A_m,\quad 
L=g M g^{-1},\\ \label{g}
g&=&{\rm diag}(e^{q_0/2},e^{q_1/2},\dots),\\ \label{M}
M&=&\left(\begin{array}{cccc}
a_0 & b_0 & 0 & \ddots  \\
b_0 & a_1 & b_1 & \ddots   \\
0 & b_1 & a_2 &  \ddots  \\
\ddots & \ddots  & \ddots & \ddots
\end{array}\right)
\eea
As a generalization of \eqref{G0} we define 
\bea
\label{Gt}
G_{nm}(t)=\langle A_n(t),A_m\rangle
\eea
and evaluate it in terms of Lanczos coefficients (we use matrix notations here for brevity)
\bea
\label{originalGt}
G(t)=g e^{M t} g^T.
\eea
The original correlation function is then $C(t)=G_{00}(t)=\langle A_0,A_0\rangle(e^{Mt})_{00}$.

As we will see now, the recursion method is closely related to classical Toda chain.
Namely Lanczos coefficients $a_n,b_n$ can be promoted to be $t$-dependent  and will satisfy Toda equations of motion. First, we interpret $\langle A(t),B\rangle$ for some operators $A,B$ as a $t$-dependent family of scalar products (Hermitian forms),
\bea
\langle A,B\rangle_t\equiv \langle A(t),B\rangle. 
\eea
It is easy to see that $\langle\, \, , \, \rangle_t$ can be defined with help of \eqref{scalarproduct} with some new $t$-dependent $\rho^t_{1,2}$,
\bea
\rho_1^t=e^{tH/2}\rho_1\, e^{tH/2},\quad \rho_2^t=e^{-tH/2}\rho_2\, e^{-tH/2}.
\eea
For any real $t$, $\rho^t_{1,2}$ satisfy the requirements we outlined for $\rho_{1,2}$ above: they are Hermitian positive semi-definite and commute with $H$. We therefore can apply the recursion method to define Krylov basis starting from the same initial $A$ for any given value of $t$. This defines the orthogonal basis $A_n^t$, $A_0^t\equiv A$,
\bea
\label{tG}
G_{nm}^t\equiv \langle A_n^t, A_m^t\rangle_t =\delta_{nm}e^{q_n(t)},
\eea 
where $a_n,b_n$ and $q_n$ are now $t$-dependent, 
\bea
\label{an}
a_n(t)&=&{\langle [H,A^t_n],A^t_n\rangle_t \over \langle A^t_n,A^t_n\rangle_t},\\
b_n^2(t)&=&e^{q_{n+1}-q_n},\quad q_n(t)=\ln\langle A_n^t,A_n^t\rangle_t.
\eea
With help of $a_n(t),b_n(t),q_n(t)$ we also define $t$-dependent matrices $M(t)$ and $g(t)$, see eqs.~(\ref{g},\ref{M}).

%Since $\rho_i$ commute with the Hamiltonian we can rewrite $G_{nm}(t)$ \eqref{Gt} as follows
%\bea
%G_{nm}(t)=\langle A_n(t/2),A_m(t/2)\rangle
%\eea
A crucial observation is that $G_{nm}(t)$ \eqref{Gt} and $G_{nm}^t$ \eqref{tG} are the matrix representation of the same scalar product $\langle \, \, ,\, \rangle_t$ written in terms of two different bases, $A_n$ and $A_n^t$. They are therefore related by a change of coordinates 
\bea
\label{maine}
G(t)=z(t)G^t z(t)^T,\\
A_n=\sum_m z_{nm}(t) A_m^t. \label{z}
\eea 
Going back to the definition \eqref{iteration}, for any given $t$, basis element $A_n^t$ is a linear combination of nested commutators 
$\underbrace{[H,\dots,[H,A]]}_{k\, \rm times}$ with $0\leq k\leq n$ such that the coefficient in front of the nested commutator of degree $n$ is exactly one. Therefore matrix $z(t)$ which transforms basis $A_n^t$ into basis $A_n\equiv A_n^{t=0}$ is lower-triangular with the identities on the diagonal. 
For convenience we rewrite \eqref{maine} using \eqref{originalGt} and express $G^t$ in terms of $g(t)$,
\bea
\label{main}
G(t)=g(0) e^{M(0)t} g(0)^T=z(t) g(t) g(t)^T z(t)^T.
\eea
The right-hand-side of \eqref{main} defines orispherical coordinate system $(q_n, z_{nm})$, $n>m$, on the space of symmetric positive-definite matrices $G$. Explicit time dependence of $G(t)$ given by \eqref{main} provides that 
\bea
{d\over dt}\left(G^{-1}\dot{G}\right)=0. \label{flow}
\eea
Thus, $G(t)$ describes a geodesic flow on the space of symmetric positive-definite matrices, which is projected onto the space of diagonal matrices (parametrized by coordinates $q_n$) by the group of lower-triangular matrices with the identities on the diagonal. This flow is described by an open Toda chain,  which was shown by applying the Hamiltonian reduction formalism toward the original geodesic flow \cite{Ol'shanetskii1980}. From here follows that $q_n(t)$ satisfy Toda equations
\bea
\ddot{q}_n=e^{q_{n+1}-q_n}-e^{q_{n}-q_{n-1}},
\eea
which can be written in Flaschka form
\bea
\label{Flaschka}
\dot{a}_n=b_n^2-b_{n-1}^2,\quad \dot{b}_n=b_n(a_{n+1}-a_n)/2.
\eea
The relation between $a_n,b_n$ and $q_n$ is given by 
\bea
a_n(t)\equiv \dot{q}_n,\quad b_n(t)&\equiv &e^{(q_{n+1}-q_n)/2},
\eea
which is consistent with \eqref{an} if we take into account that the derivative of $A_n^t$ with respect to $t$ is a linear combination of $A_k^t$ for $0\leq k<n$.

Alternatively, Toda equations can be written in Hirota's bilinear form 
\bea
\tau_n \ddot{\tau}_n-\dot{\tau_n}^2=\tau_{n+1}\tau_{n-1}, \quad \tau_{-1}\equiv 1,
\eea
where $\tau_n={\rm exp}(\sum_{0\leq k\leq n} q_n)$ are the leading principal minors of $G_{nm}(t)$. In particular 
\bea
\tau_0(t)=e^{q_0(t)}=C(t),
\eea
which establishes in full generality that time-correlation function analytically continued to Euclidean time is a tau-function of Toda hierarchy. 

By virtue of the identity $\langle A(t/2),B(t/2)\rangle=\langle A,B\rangle_t$ for any $A,B$ the operators $A_n^t(t/2)$ define the orthogonal Krylov basis associated with the initial scalar product $\langle \, \, ,\, \rangle$ and the initial operator $A(t/2)$. Corresponding Lanczos coefficients are $a_n(t)$ and $b_n(t)$.  This follows from the fact that the relation \eqref{iteration} is linear, and hence will hold if all operators are evolved in time. Thus, the flow described by the Toda chain can be defined solely in terms of the original scalar product, by considering different initial vectors of the Krylov basis. Furthermore, since $e^{-q_n(0)/2}A_n$ and $e^{-q_n(t)/2}A_n^t(t/2)$ are orthonormal bases for the same scalar product, they must be related by an orthogonal transformation $Q^T$,
\bea
\label{Q}
\sum_m Q^T_{nm}(t/2)e^{-q_m(0)/2}A_m=e^{-q_n(t)/2}A_n^t(t/2).
\eea 
Evolving this equation in time by $-t/2$ and using the relation \eqref{z} between $A_n$ and $A_n^t$  we find
\bea
\label{QR}
e^{M(0)t}=Q(t)R(t),\quad R^T(t/2)=g(0)^{-1} z(t)g(t).\ \ 
\eea
This defines ``QR'' decomposition of  $e^{M(0)t}$  \cite{symes1980hamiltonian}. 

We did not require the Hermitian form $\langle \, \, ,\, \rangle$ to be positive-definite, merely positive semi-definite. Therefore the coefficient $b_n^2$ given by  \eqref{ab} may vanish either because $A_{n+1}$ vanishes as an operator, or because it has zero ``norm'' $\langle A_{n+1},A_{n+1}\rangle=0$. In either case time-evolved $A(t)$ will be a linear combination  of only first $n$ basis elements $A_k$, and therefore $C(t)$ will be described in terms of a finite Toda chain. Matrix $G_{kl}$ in this case will be defined for $0\leq k,l\leq n$ and will be finite  positive-definite.  Thus, without loss of generality matrix $G$ is always positive-definite, which justifies taking inverse in \eqref{flow}.  This completes the construction of the recursion method as a part of the Toda chain flow in Krylov space.

A few comments are in order. The construction above is linear in  scalar product, and therefore applies to any linear combination of \eqref{scalarproduct}, i.e.~when 
\bea
\langle A, B\rangle\equiv \sum_i {\rm Tr}(A^\dagger \rho^{(i)}_1 B \rho^{(i)}_2\rangle.
\eea
This may appear e.g. in the context of differently order thermal correlators. For example if $\rho_1=\rho_2=\rho^{1/2}, \rho=e^{-\beta H}/Z$, this defines symmetric ordering with $a_n=0$. Conventional thermal correlator is obtained by $\rho_1={\mathcal I}$, $\rho_2=\rho$. In this case Lanczos coefficients are related to those in the symmetric case by time-evolving them using Toda equations of motion from $t=0$ to $t=\beta/2$. 
%There is also a third choice, $\rho^{(1)}_1={\mathcal I}/2$, $\rho^{(1)}_2=\rho$, and $\rho^{(2)}_1=\rho$, $\rho^{(2)}_2={\mathcal I}/2$, which can be obtained from the previous one by symmetrization $C(t)\rightarrow (C(t)+C(-t))/2$.% In this case corresponding Lanczos coefficients can be obtained from $C(t)$ using e.g.~Hankel representation introduced below. 

Finally, we remark that the relation between the recursion method and Toda chain is not limited to  time-correlation function, and can be readily extended to other cases whenever the recursion method applies. Furthermore, continued fraction representation of the Green's function appears naturally  in the context of Toda chain dynamics. This is explained in supplemental materials. 

The relation to Toda chain  provides a new way to analyze the time-correlation function. Below we apply it to elucidate chaos in quantum many-body systems. The growth of $C(t)$ in Euclidean time is qualitatively different in integrable (solvable) and generic lattice systems \cite{alex2019euclidean}. Considering thermodynamic limit, in known integrable examples $C(t)$ is an entire function of a complex parameter $t$. 
On the contrary, an accurate counting of nested commutators appearing in the Taylor series expansion of $C(t)$ suggests that in general, i.e.~non-integrable case, $C(t)$ will be singular at some finite $t=t^*$. 
This behavior is confirmed by an explicit example of \cite{Bouch}. The same singular behavior for the chaotic models follows from the conjecture of \cite{Parker_2019}, which associates chaos with the maximal rate of growth of Lanczos coefficients, $b_n \propto n$, permitted by analyticity of $C(t)$ at $t=0$. An equivalent formulation in terms of the  power spectrum of $C(t)$ was advocated earlier in \cite{elsayed2014signatures}.

The original analysis of \cite{Parker_2019} assumed $a_n=0$.
The Toda chain formalism provides an easy way to extend this result. From the equations of motion \eqref{Flaschka} it follow that linear growth $b_n \propto n$ is consistent with at most linear growth of $a_n$ and the slope of $a_n$ can not exceed twice the slope of $b_n$. 
This can be illustrated with the  help of a family of exact solutions. Combining \eqref{Flaschka} into 
\bea
\label{eomb}
{d^2\over dt^2}\ln b_n^2=b_{n+1}^2-2b_n+b_{n-1}^2,
\eea
and assuming $b^2_n=b^2(t)p(n)$ where $p(n)$ is an arbitrary quadratic polynomial, we find a family of solutions,
\bea
\label{exacts}
a_n(t)&=&(2n+c)J \cot (J(t_0-t)),\\
b^2_n(t)&=&{ (n+c)(n+1) J^2 \over \sin^{2}(J(t_0-t))}.
\eea 
This family is associated with the tau-function $\tau_0 \propto (\sin(J(t_0-t)))^{-c}$,
%\bea
%C(t)=\tau_0={\sin(J(t_0-t))\over \sin(J(t_0-t))}
%\eea
which is the time-correlation function of the SYK model  \cite{PhysRevD.94.106002,Parker_2019}. The same solution with $c=2$ in the $J\rightarrow 0$ limit also appeared in \cite{Grassi:2019txd} in the context of ${\mathcal N}=2$ SYM. At large $n$, $a_n/b_n \propto 2 \sin(J(t_0-t))$. Thus, in generality, chaotic behavior is reflected by the linear growth of both $a_n$ and $b_n$, parametrized by $J$ and dimensionless $|\gamma|\leq 1$, 
\bea
\lim_n\,  (b_n^2-a_n^2/4)/n^2= J^2, \quad 
\lim_n\,   a_n/b_n= 2\gamma.
\eea
The asymptotic behavior of $a_n,b_n$ controls the  location of the singularity of $C(t)$, $t^*=\arcsin(\gamma)/J$.

Singular behavior of the time-correlation function can be further elucidated. As a starting point we assume that $C(t)=G_{00}(t)$ is a smooth function, together with its derivatives for $0\leq t<t^*$ and diverges at $t=t^*$.
From here follows that for $n,m\geq 0$, $G_{nm}(t)$ defined in \eqref{Gt}  is regular for $0\leq t<t^*$. Indeed, different   matrix elements $G_{nm}(t)$ are related by the differential operator
\bea
\label{iterationdt}
G_{n+1,m}=\left({d\over dt}-a_n\right)G_{nm}-b_{n-1}^2 G_{n-1,m}.
\eea
Therefore all $G_{nm}$ are regular for $0\leq t<t^*$, provided $C(t)$ is sufficiently smooth.

Using QR decomposition \eqref{QR} we can decompose
\bea
\label{R00}
R_{00}(t/2)^2=C(t)/C(0), 
\eea
and conclude that $R_{00}(t)$ is regular for $0\leq t<t^*/2$ and diverges at $t=t^*/2$.  
Using  \eqref{QR} again we can decompose $A(t)$ into orthonormal Krylov basis
\bea
e^{-q_0(0)/2}A(t)=\sum_n c_n(t) \left(e^{-q_n(0)/2}A_n\right),
\eea
where 
\bea
\label{cn}
c_n(t)\equiv e^{-(q_0(0)+q_n(0))/2} G_{0n}(t)=R_{00}(t) Q_{n0}(t).
\eea
Here $R_{00}(t)$ is the norm of the operator and unit vector $Q_{n0}(t)$ specifies projection on a particular basis element. 
Regularity of $G_{0n}(t)$ at $t=t^*/2$ and divergence of $R_{00}(t)$ at  $t=t^*/2$ implies $Q_{n0}(t^*/2)$ for all $n$ have to vanish.  This is a manifestation of delocalization in Krylov space:  at $t=t^*/2$ the operator $A(t)$ spreads across the whole Krylov space, such that its norm diverges, while its projection on any particular normalized basis element is finite. The same can be seen from the inverse participation ratio $I$, 
\bea
\label{IPR}
I\equiv \left(\sum_n Q_{n0}^4\right)^{-1},
\eea
which diverges at $t=t^*/2$.
We illustrate this behavior using explicit solution \eqref{exacts} and show that it correctly captures universal behavior near $t=t^*/2$  in supplemental materials.  

Starting from $b_n^2=b^2 p(n)$ where $p(n)$ is a linear function one finds an explicit solution illustrating ``integrable'' behavior $b_n\propto n^{1/2}$. The corresponding tau-function grows double-exponentially $\tau_0\propto {\rm exp}\{e^{m(t-t_0)}\}$ which is the behavior of $C(t)$ in generic  one-dimensional systems \cite{alex2019euclidean}. This further emphasizes that non-integrable one-dimensional systems can not be considered fully chaotic.  In the limit $m\rightarrow 0$ tau-function becomes Gaussian, $\tau_0\propto e^{a(t-t_0)^2/2}$. In both cases $A(t)$ is moving as a localized wave-packet in Krylov space, with the inverse participation ratio growing with time exponentially when $m\neq 0$ or merely linearly when $\tau_0$ is a Gaussian. Technical details can be found in supplemental materials.

{\it Discussion.} 
In this Letter we established an explicit representation for the Euclidean time evolution of the time-correlation function as classical dynamics of  integrable Toda chain. We have subsequently used the Toda chain formalism to elucidate the behavior of time-correlation function in non-integrable quantum many-body systems. We have extended the conjecture of \cite{Parker_2019}. to include non-vanishing $a_n$. We have also  demonstrated that singularity along the imaginary time axis, which is a generic behavior for non-integrable systems, is due to delocalization in Krylov space. 

The connection between the recursion method and Toda chain is likely to lead to new practical improvements in numerical applications, as suggested by many other uses of Toda chain in the context of  computational algorithms \cite{nakamura2004new}.

Tau-functions of completely integrable systems  have free fermion representation \cite{jimbo1983solitons}. It is a natural question to ask how this representation may appear in the context of the time-correlation function of a {\it generic} Hamiltonian system. 
The construction  presented in this Letter does not require  the system to be quantum.  In the classical case scalar product \eqref{scalarproduct} can be defined as an integral over the phase space, and the adjoint action $[H,\, \, \, ]$ in \eqref{iteration} will be substituted  by the Poisson brackets.  Further, an arbitrary classical system can be reformulated in terms of supersymmetric path integral, which includes auxiliary fermionic degrees of freedom \cite{Gozzi:1993tm}. We expect free fermion representation to follow from here. 

%Additional this formalism  
%However one could wonder what is the proper substitute of the Krylov basis in the Hilbert space.
%Conjecturally  the reformulation of an arbitrary classical system  in terms of a path integral \cite{Gozzi:1993tm} provides the proper framework. In this approach the supersymmetric  quantum mechanics emerges and the Liovillian was identified with the bosonic part of the Hamiltonian. The Hilbert space can be identified and presumably the fermionic degrees of freedom involved in this path integral representation fit with the fermionic representation of the Toda chain.

In retrospect appearance of Toda chain in the context of the recursion method is not that surprising. Completely integrable dynamics often can be understood in terms of group theory, as a free motion on  a symmetric space \cite{olper}. In this Letter we developed this picture for the  time-correlation function and  identified free motion with the rotation of basis in Krylov space. There is a more general group-theoretic framework behind  integrable dynamics  emerging in the context of a generic non-integrable system. The corresponding group is the group of canonical transformations specified by the phase space of the original system. In this framework non-integrable Hamiltonian of the original system defines a particular solution of integrable dynamics. There is a number of explicit examples \cite{gorsky,aganagic,Bettelheim_2006} when this construction was elucidated. We leave for the future the task of understanding our results from this point of view. 

% 
% gives rise 
%to the group of canonical transformations, 
%key concerns the  group of the canonical transformations of the phase spaces themselves. The choice of the Hamiltonian provides the choice of the solution in the corresponding integrable model.
%The examples of the emergence of the integrable dynamics from a dynamics on the group of canonical transformations  are provided in the context of topological strings \cite{aganagic}  for holomorphic phase space variables and for real phase space variables in \cite{gorsky,Bettelheim_2006}.

One of our main results is the relation between non-integrability of the original physical system and delocalization in Krylov space. This results can be understood in the context of a general idea that localization or ergodicity in physical space corresponds to localization or delocalization in the auxiliary ``Fock space'' of  a ``particle'' moving on a  graph \cite{altshuler1997quasiparticle}. This idea has been further developed in the context of many-body localization in \cite{basko2006metal}. In a general case construction of the appropriate graph is not clear. Our study suggests that Krylov basis provides a suitable representation of the ``Fock space,''  with the tridiagonal Liouvillian matrix $M$ describing hoping of a particle on a one-dimensional graph.

\acknowledgments
We thank Alex Avdoshkin, Boris Fine, and Zohar Komargodski for discussions. 
The work of A.G. was supported
by Basis Foundation fellowship and RFBR grant 19-02-00214. 
The authors thank Simons Center for Geometry and Physics at Stony Brook
University, at which some or this research was performed, for the hospitality and support.

\bibliography{Letter}

\begin{thebibliography}{27}%
\makeatletter
\providecommand \@ifxundefined [1]{%
 \@ifx{#1\undefined}
}%
\providecommand \@ifnum [1]{%
 \ifnum #1\expandafter \@firstoftwo
 \else \expandafter \@secondoftwo
 \fi
}%
\providecommand \@ifx [1]{%
 \ifx #1\expandafter \@firstoftwo
 \else \expandafter \@secondoftwo
 \fi
}%
\providecommand \natexlab [1]{#1}%
\providecommand \enquote  [1]{``#1''}%
\providecommand \bibnamefont  [1]{#1}%
\providecommand \bibfnamefont [1]{#1}%
\providecommand \citenamefont [1]{#1}%
\providecommand \href@noop [0]{\@secondoftwo}%
\providecommand \href [0]{\begingroup \@sanitize@url \@href}%
\providecommand \@href[1]{\@@startlink{#1}\@@href}%
\providecommand \@@href[1]{\endgroup#1\@@endlink}%
\providecommand \@sanitize@url [0]{\catcode `\\12\catcode `\$12\catcode
  `\&12\catcode `\#12\catcode `\^12\catcode `\_12\catcode `\%12\relax}%
\providecommand \@@startlink[1]{}%
\providecommand \@@endlink[0]{}%
\providecommand \url  [0]{\begingroup\@sanitize@url \@url }%
\providecommand \@url [1]{\endgroup\@href {#1}{\urlprefix }}%
\providecommand \urlprefix  [0]{URL }%
\providecommand \Eprint [0]{\href }%
\providecommand \doibase [0]{http://dx.doi.org/}%
\providecommand \selectlanguage [0]{\@gobble}%
\providecommand \bibinfo  [0]{\@secondoftwo}%
\providecommand \bibfield  [0]{\@secondoftwo}%
\providecommand \translation [1]{[#1]}%
\providecommand \BibitemOpen [0]{}%
\providecommand \bibitemStop [0]{}%
\providecommand \bibitemNoStop [0]{.\EOS\space}%
\providecommand \EOS [0]{\spacefactor3000\relax}%
\providecommand \BibitemShut  [1]{\csname bibitem#1\endcsname}%
\let\auto@bib@innerbib\@empty
%</preamble>
\bibitem [{\citenamefont
  {Cyrot-Lackmann}(1967)}]{doi:10.1080/00018736700101495}%
  \BibitemOpen
  \bibfield  {author} {\bibinfo {author} {\bibfnamefont {F.}~\bibnamefont
  {Cyrot-Lackmann}},\ }\bibfield  {title} {\enquote {\bibinfo {title} {On the
  electronic structure of liquid transitional metals},}\ }\href {\doibase
  10.1080/00018736700101495} {\bibfield  {journal} {\bibinfo  {journal}
  {Advances in Physics}\ }\textbf {\bibinfo {volume} {16}},\ \bibinfo {pages}
  {393--400} (\bibinfo {year} {1967})},\ \Eprint
  {http://arxiv.org/abs/https://doi.org/10.1080/00018736700101495}
  {https://doi.org/10.1080/00018736700101495} \BibitemShut {NoStop}%
\bibitem [{\citenamefont {Mori}(1965)}]{mori1965continued}%
  \BibitemOpen
  \bibfield  {author} {\bibinfo {author} {\bibfnamefont {Hazime}\ \bibnamefont
  {Mori}},\ }\bibfield  {title} {\enquote {\bibinfo {title} {A
  continued-fraction representation of the time-correlation functions},}\
  }\href@noop {} {\bibfield  {journal} {\bibinfo  {journal} {Progress of
  Theoretical Physics}\ }\textbf {\bibinfo {volume} {34}},\ \bibinfo {pages}
  {399--416} (\bibinfo {year} {1965})}\BibitemShut {NoStop}%
\bibitem [{\citenamefont {Haydock}(1980)}]{haydock1980recursive}%
  \BibitemOpen
  \bibfield  {author} {\bibinfo {author} {\bibfnamefont {Roger}\ \bibnamefont
  {Haydock}},\ }\bibfield  {title} {\enquote {\bibinfo {title} {The recursive
  solution of the schr{\"o}dinger equation},}\ }\href@noop {} {\bibfield
  {journal} {\bibinfo  {journal} {Computer Physics Communications}\ }\textbf
  {\bibinfo {volume} {20}},\ \bibinfo {pages} {11--16} (\bibinfo {year}
  {1980})}\BibitemShut {NoStop}%
\bibitem [{\citenamefont {Izergin}\ and\ \citenamefont
  {Korepin}(1984)}]{izergin1984quantum}%
  \BibitemOpen
  \bibfield  {author} {\bibinfo {author} {\bibfnamefont {AG}~\bibnamefont
  {Izergin}}\ and\ \bibinfo {author} {\bibfnamefont {VE}~\bibnamefont
  {Korepin}},\ }\bibfield  {title} {\enquote {\bibinfo {title} {The quantum
  inverse scattering method approach to correlation functions},}\ }\href@noop
  {} {\bibfield  {journal} {\bibinfo  {journal} {Communications in Mathematical
  Physics}\ }\textbf {\bibinfo {volume} {94}},\ \bibinfo {pages} {67--92}
  (\bibinfo {year} {1984})}\BibitemShut {NoStop}%
\bibitem [{\citenamefont {Bettelheim}\ \emph {et~al.}(2006)\citenamefont
  {Bettelheim}, \citenamefont {Abanov},\ and\ \citenamefont
  {Wiegmann}}]{Bettelheim_2006}%
  \BibitemOpen
  \bibfield  {author} {\bibinfo {author} {\bibfnamefont {E.}~\bibnamefont
  {Bettelheim}}, \bibinfo {author} {\bibfnamefont {A.~G.}\ \bibnamefont
  {Abanov}}, \ and\ \bibinfo {author} {\bibfnamefont {P.}~\bibnamefont
  {Wiegmann}},\ }\bibfield  {title} {\enquote {\bibinfo {title} {Orthogonality
  catastrophe and shock waves in a nonequilibrium fermi gas},}\ }\href
  {\doibase 10.1103/physrevlett.97.246402} {\bibfield  {journal} {\bibinfo
  {journal} {Physical Review Letters}\ }\textbf {\bibinfo {volume} {97}}
  (\bibinfo {year} {2006}),\ 10.1103/physrevlett.97.246402}\BibitemShut
  {NoStop}%
\bibitem [{\citenamefont {Gorsky}\ \emph {et~al.}(1998)\citenamefont {Gorsky},
  \citenamefont {Marshakov}, \citenamefont {Mironov},\ and\ \citenamefont
  {Morozov}}]{gmmm}%
  \BibitemOpen
  \bibfield  {author} {\bibinfo {author} {\bibfnamefont {A.}~\bibnamefont
  {Gorsky}}, \bibinfo {author} {\bibfnamefont {A.}~\bibnamefont {Marshakov}},
  \bibinfo {author} {\bibfnamefont {A.}~\bibnamefont {Mironov}}, \ and\
  \bibinfo {author} {\bibfnamefont {A.}~\bibnamefont {Morozov}},\ }\bibfield
  {title} {\enquote {\bibinfo {title} {{RG equations from Whitham
  hierarchy}},}\ }\href {\doibase 10.1016/S0550-3213(98)00315-0} {\bibfield
  {journal} {\bibinfo  {journal} {Nucl. Phys.}\ }\textbf {\bibinfo {volume}
  {B527}},\ \bibinfo {pages} {690--716} (\bibinfo {year} {1998})},\ \Eprint
  {http://arxiv.org/abs/hep-th/9802007} {arXiv:hep-th/9802007 [hep-th]}
  \BibitemShut {NoStop}%
%%CITATION = HEP-TH/9802007;%%
\bibitem [{\citenamefont {Grassi}\ \emph {et~al.}(2019)\citenamefont {Grassi},
  \citenamefont {Komargodski},\ and\ \citenamefont {Tizzano}}]{Grassi:2019txd}%
  \BibitemOpen
  \bibfield  {author} {\bibinfo {author} {\bibfnamefont {Alba}\ \bibnamefont
  {Grassi}}, \bibinfo {author} {\bibfnamefont {Zohar}\ \bibnamefont
  {Komargodski}}, \ and\ \bibinfo {author} {\bibfnamefont {Luigi}\ \bibnamefont
  {Tizzano}},\ }\bibfield  {title} {\enquote {\bibinfo {title} {{Extremal
  Correlators and Random Matrix Theory}},}\ }\href@noop {} {\  (\bibinfo {year}
  {2019})},\ \Eprint {http://arxiv.org/abs/1908.10306} {arXiv:1908.10306
  [hep-th]} \BibitemShut {NoStop}%
%%CITATION = ARXIV:1908.10306;%%
\bibitem [{\citenamefont {Shenker}\ and\ \citenamefont
  {Stanford}(2014)}]{shenker2014black}%
  \BibitemOpen
  \bibfield  {author} {\bibinfo {author} {\bibfnamefont {Stephen~H}\
  \bibnamefont {Shenker}}\ and\ \bibinfo {author} {\bibfnamefont {Douglas}\
  \bibnamefont {Stanford}},\ }\bibfield  {title} {\enquote {\bibinfo {title}
  {Black holes and the butterfly effect},}\ }\href@noop {} {\bibfield
  {journal} {\bibinfo  {journal} {Journal of High Energy Physics}\ }\textbf
  {\bibinfo {volume} {2014}},\ \bibinfo {pages} {67} (\bibinfo {year}
  {2014})}\BibitemShut {NoStop}%
\bibitem [{\citenamefont {D'Alessio}\ \emph {et~al.}(2016)\citenamefont
  {D'Alessio}, \citenamefont {Kafri}, \citenamefont {Polkovnikov},\ and\
  \citenamefont {Rigol}}]{doi:10.1080/00018732.2016.1198134}%
  \BibitemOpen
  \bibfield  {author} {\bibinfo {author} {\bibfnamefont {Luca}\ \bibnamefont
  {D'Alessio}}, \bibinfo {author} {\bibfnamefont {Yariv}\ \bibnamefont
  {Kafri}}, \bibinfo {author} {\bibfnamefont {Anatoli}\ \bibnamefont
  {Polkovnikov}}, \ and\ \bibinfo {author} {\bibfnamefont {Marcos}\
  \bibnamefont {Rigol}},\ }\bibfield  {title} {\enquote {\bibinfo {title} {From
  quantum chaos and eigenstate thermalization to statistical mechanics and
  thermodynamics},}\ }\href {\doibase 10.1080/00018732.2016.1198134} {\bibfield
   {journal} {\bibinfo  {journal} {Advances in Physics}\ }\textbf {\bibinfo
  {volume} {65}},\ \bibinfo {pages} {239--362} (\bibinfo {year} {2016})},\
  \Eprint {http://arxiv.org/abs/https://doi.org/10.1080/00018732.2016.1198134}
  {https://doi.org/10.1080/00018732.2016.1198134} \BibitemShut {NoStop}%
\bibitem [{\citenamefont {Parker}\ \emph {et~al.}(2019)\citenamefont {Parker},
  \citenamefont {Cao}, \citenamefont {Avdoshkin}, \citenamefont {Scaffidi},\
  and\ \citenamefont {Altman}}]{Parker_2019}%
  \BibitemOpen
  \bibfield  {author} {\bibinfo {author} {\bibfnamefont {Daniel~E.}\
  \bibnamefont {Parker}}, \bibinfo {author} {\bibfnamefont {Xiangyu}\
  \bibnamefont {Cao}}, \bibinfo {author} {\bibfnamefont {Alexander}\
  \bibnamefont {Avdoshkin}}, \bibinfo {author} {\bibfnamefont {Thomas}\
  \bibnamefont {Scaffidi}}, \ and\ \bibinfo {author} {\bibfnamefont {Ehud}\
  \bibnamefont {Altman}},\ }\bibfield  {title} {\enquote {\bibinfo {title} {A
  universal operator growth hypothesis},}\ }\href {\doibase
  10.1103/physrevx.9.041017} {\bibfield  {journal} {\bibinfo  {journal}
  {Physical Review X}\ }\textbf {\bibinfo {volume} {9}} (\bibinfo {year}
  {2019}),\ 10.1103/physrevx.9.041017}\BibitemShut {NoStop}%
\bibitem [{\citenamefont {Avdoshkin}\ and\ \citenamefont
  {Dymarsky}(2019)}]{alex2019euclidean}%
  \BibitemOpen
  \bibfield  {author} {\bibinfo {author} {\bibfnamefont {Alexander}\
  \bibnamefont {Avdoshkin}}\ and\ \bibinfo {author} {\bibfnamefont {Anatoly}\
  \bibnamefont {Dymarsky}},\ }\href@noop {} {\enquote {\bibinfo {title}
  {Euclidean operator growth and quantum chaos},}\ } (\bibinfo {year} {2019}),\
  \Eprint {http://arxiv.org/abs/1911.09672} {arXiv:1911.09672
  [cond-mat.stat-mech]} \BibitemShut {NoStop}%
\bibitem [{\citenamefont {Grigolini}\ \emph {et~al.}(1983)\citenamefont
  {Grigolini}, \citenamefont {Grosso}, \citenamefont {Parravicini},\ and\
  \citenamefont {Sparpaglione}}]{PhysRevB.27.7342}%
  \BibitemOpen
  \bibfield  {author} {\bibinfo {author} {\bibfnamefont {P.}~\bibnamefont
  {Grigolini}}, \bibinfo {author} {\bibfnamefont {G.}~\bibnamefont {Grosso}},
  \bibinfo {author} {\bibfnamefont {G.~Pastori}\ \bibnamefont {Parravicini}}, \
  and\ \bibinfo {author} {\bibfnamefont {M.}~\bibnamefont {Sparpaglione}},\
  }\bibfield  {title} {\enquote {\bibinfo {title} {Calculation of relaxation
  functions: A new development within the mori formalism},}\ }\href {\doibase
  10.1103/PhysRevB.27.7342} {\bibfield  {journal} {\bibinfo  {journal} {Phys.
  Rev. B}\ }\textbf {\bibinfo {volume} {27}},\ \bibinfo {pages} {7342--7347}
  (\bibinfo {year} {1983})}\BibitemShut {NoStop}%
\bibitem [{\citenamefont {Viswanath}\ and\ \citenamefont
  {M{\"u}ller}(2008)}]{viswanath2008recursion}%
  \BibitemOpen
  \bibfield  {author} {\bibinfo {author} {\bibfnamefont {VS}~\bibnamefont
  {Viswanath}}\ and\ \bibinfo {author} {\bibfnamefont {Gerhard}\ \bibnamefont
  {M{\"u}ller}},\ }\href@noop {} {\emph {\bibinfo {title} {The Recursion
  Method: Application to Many-Body Dynamics}}},\ Vol.~\bibinfo {volume} {23}\
  (\bibinfo  {publisher} {Springer Science \& Business Media},\ \bibinfo {year}
  {2008})\BibitemShut {NoStop}%
\bibitem [{\citenamefont {Ol'shanetskii}\ and\ \citenamefont
  {Perelomov}(1980)}]{Ol'shanetskii1980}%
  \BibitemOpen
  \bibfield  {author} {\bibinfo {author} {\bibfnamefont {M.~A.}\ \bibnamefont
  {Ol'shanetskii}}\ and\ \bibinfo {author} {\bibfnamefont {A.~M.}\ \bibnamefont
  {Perelomov}},\ }\bibfield  {title} {\enquote {\bibinfo {title} {The toda
  chain as a reduced system},}\ }\href {\doibase 10.1007/BF01047139} {\bibfield
   {journal} {\bibinfo  {journal} {Theoretical and Mathematical Physics}\
  }\textbf {\bibinfo {volume} {45}},\ \bibinfo {pages} {843--854} (\bibinfo
  {year} {1980})}\BibitemShut {NoStop}%
\bibitem [{\citenamefont {Symes}(1980)}]{symes1980hamiltonian}%
  \BibitemOpen
  \bibfield  {author} {\bibinfo {author} {\bibfnamefont {WW}~\bibnamefont
  {Symes}},\ }\bibfield  {title} {\enquote {\bibinfo {title} {Hamiltonian group
  actions and integrable systems},}\ }\href@noop {} {\bibfield  {journal}
  {\bibinfo  {journal} {Physica D: Nonlinear Phenomena}\ }\textbf {\bibinfo
  {volume} {1}},\ \bibinfo {pages} {339--374} (\bibinfo {year}
  {1980})}\BibitemShut {NoStop}%
\bibitem [{\citenamefont {Bouch}(2015)}]{Bouch}%
  \BibitemOpen
  \bibfield  {author} {\bibinfo {author} {\bibfnamefont {Gabeiel}\ \bibnamefont
  {Bouch}},\ }\bibfield  {title} {\enquote {\bibinfo {title} {{Complex-time
  singularity and locality estimates for quantum lattice systems}},}\ }\href
  {\doibase 10.1063/1.4936209} {\bibfield  {journal} {\bibinfo  {journal}
  {Journal of Mathematical Physics}\ }\textbf {\bibinfo {volume} {56}},\
  \bibinfo {pages} {123303} (\bibinfo {year} {2015})}\BibitemShut {NoStop}%
\bibitem [{\citenamefont {Elsayed}\ \emph {et~al.}(2014)\citenamefont
  {Elsayed}, \citenamefont {Hess},\ and\ \citenamefont
  {Fine}}]{elsayed2014signatures}%
  \BibitemOpen
  \bibfield  {author} {\bibinfo {author} {\bibfnamefont {Tarek~A}\ \bibnamefont
  {Elsayed}}, \bibinfo {author} {\bibfnamefont {Benjamin}\ \bibnamefont
  {Hess}}, \ and\ \bibinfo {author} {\bibfnamefont {Boris~V}\ \bibnamefont
  {Fine}},\ }\bibfield  {title} {\enquote {\bibinfo {title} {Signatures of
  chaos in time series generated by many-spin systems at high temperatures},}\
  }\href@noop {} {\bibfield  {journal} {\bibinfo  {journal} {Physical Review
  E}\ }\textbf {\bibinfo {volume} {90}},\ \bibinfo {pages} {022910} (\bibinfo
  {year} {2014})}\BibitemShut {NoStop}%
\bibitem [{\citenamefont {Maldacena}\ and\ \citenamefont
  {Stanford}(2016)}]{PhysRevD.94.106002}%
  \BibitemOpen
  \bibfield  {author} {\bibinfo {author} {\bibfnamefont {Juan}\ \bibnamefont
  {Maldacena}}\ and\ \bibinfo {author} {\bibfnamefont {Douglas}\ \bibnamefont
  {Stanford}},\ }\bibfield  {title} {\enquote {\bibinfo {title} {Remarks on the
  sachdev-ye-kitaev model},}\ }\href {\doibase 10.1103/PhysRevD.94.106002}
  {\bibfield  {journal} {\bibinfo  {journal} {Phys. Rev. D}\ }\textbf {\bibinfo
  {volume} {94}},\ \bibinfo {pages} {106002} (\bibinfo {year}
  {2016})}\BibitemShut {NoStop}%
\bibitem [{\citenamefont {Nakamura}(2004)}]{nakamura2004new}%
  \BibitemOpen
  \bibfield  {author} {\bibinfo {author} {\bibfnamefont {Yoshimasa}\
  \bibnamefont {Nakamura}},\ }\bibfield  {title} {\enquote {\bibinfo {title} {A
  new approach to numerical algorithms in terms of integrable systems},}\ }in\
  \href@noop {} {\emph {\bibinfo {booktitle} {International Conference on
  Informatics Research for Development of Knowledge Society Infrastructure,
  2004. ICKS 2004.}}}\ (\bibinfo {organization} {IEEE},\ \bibinfo {year}
  {2004})\ pp.\ \bibinfo {pages} {194--205}\BibitemShut {NoStop}%
\bibitem [{\citenamefont {Jimbo}\ and\ \citenamefont
  {Miwa}(1983)}]{jimbo1983solitons}%
  \BibitemOpen
  \bibfield  {author} {\bibinfo {author} {\bibfnamefont {Michio}\ \bibnamefont
  {Jimbo}}\ and\ \bibinfo {author} {\bibfnamefont {Tetsuji}\ \bibnamefont
  {Miwa}},\ }\bibfield  {title} {\enquote {\bibinfo {title} {Solitons and
  infinite dimensional lie algebras},}\ }\href@noop {} {\bibfield  {journal}
  {\bibinfo  {journal} {Publications of the Research Institute for Mathematical
  Sciences}\ }\textbf {\bibinfo {volume} {19}},\ \bibinfo {pages} {943--1001}
  (\bibinfo {year} {1983})}\BibitemShut {NoStop}%
\bibitem [{\citenamefont {Gozzi}\ and\ \citenamefont
  {Reuter}(1994)}]{Gozzi:1993tm}%
  \BibitemOpen
  \bibfield  {author} {\bibinfo {author} {\bibfnamefont {E.}~\bibnamefont
  {Gozzi}}\ and\ \bibinfo {author} {\bibfnamefont {M.}~\bibnamefont {Reuter}},\
  }\bibfield  {title} {\enquote {\bibinfo {title} {{Lyapunov exponents, path
  integrals and forms}},}\ }\href {\doibase 10.1016/0960-0779(94)90026-4}
  {\bibfield  {journal} {\bibinfo  {journal} {Chaos Solitons Fractals}\
  }\textbf {\bibinfo {volume} {4}},\ \bibinfo {pages} {1117--1139} (\bibinfo
  {year} {1994})}\BibitemShut {NoStop}%
%%CITATION = CSFOE,4,1117;%%
\bibitem [{\citenamefont {Olshanetsky}\ and\ \citenamefont
  {Perelomov}(1981)}]{olper}%
  \BibitemOpen
  \bibfield  {author} {\bibinfo {author} {\bibfnamefont {M.~A.}\ \bibnamefont
  {Olshanetsky}}\ and\ \bibinfo {author} {\bibfnamefont {A.~M.}\ \bibnamefont
  {Perelomov}},\ }\bibfield  {title} {\enquote {\bibinfo {title} {{Classical
  integrable finite dimensional systems related to Lie algebras}},}\ }\href
  {\doibase 10.1016/0370-1573(81)90023-5} {\bibfield  {journal} {\bibinfo
  {journal} {Phys. Rept.}\ }\textbf {\bibinfo {volume} {71}},\ \bibinfo {pages}
  {313} (\bibinfo {year} {1981})}\BibitemShut {NoStop}%
%%CITATION = PRPLC,71,313;%%
\bibitem [{\citenamefont {Gorsky}(2001)}]{gorsky}%
  \BibitemOpen
  \bibfield  {author} {\bibinfo {author} {\bibfnamefont {Alexander~S.}\
  \bibnamefont {Gorsky}},\ }\bibfield  {title} {\enquote {\bibinfo {title}
  {{Renormalization group flows on the phase spaces and tau functions for the
  generic Hamiltonian systems}},}\ }\href {\doibase
  10.1016/S0370-2693(01)00319-7, 10.1016/S0370-2693(01)00003-X} {\bibfield
  {journal} {\bibinfo  {journal} {Phys. Lett.}\ }\textbf {\bibinfo {volume}
  {B498}},\ \bibinfo {pages} {211--217} (\bibinfo {year} {2001})},\ \bibinfo
  {note} {[Erratum: Phys. Lett.B504,362(2001)]},\ \Eprint
  {http://arxiv.org/abs/hep-th/0010068} {arXiv:hep-th/0010068 [hep-th]}
  \BibitemShut {NoStop}%
%%CITATION = HEP-TH/0010068;%%
\bibitem [{\citenamefont {Aganagic}\ \emph {et~al.}(2006)\citenamefont
  {Aganagic}, \citenamefont {Dijkgraaf}, \citenamefont {Klemm}, \citenamefont
  {Marino},\ and\ \citenamefont {Vafa}}]{aganagic}%
  \BibitemOpen
  \bibfield  {author} {\bibinfo {author} {\bibfnamefont {Mina}\ \bibnamefont
  {Aganagic}}, \bibinfo {author} {\bibfnamefont {Robbert}\ \bibnamefont
  {Dijkgraaf}}, \bibinfo {author} {\bibfnamefont {Albrecht}\ \bibnamefont
  {Klemm}}, \bibinfo {author} {\bibfnamefont {Marcos}\ \bibnamefont {Marino}},
  \ and\ \bibinfo {author} {\bibfnamefont {Cumrun}\ \bibnamefont {Vafa}},\
  }\bibfield  {title} {\enquote {\bibinfo {title} {{Topological strings and
  integrable hierarchies}},}\ }\href {\doibase 10.1007/s00220-005-1448-9}
  {\bibfield  {journal} {\bibinfo  {journal} {Commun. Math. Phys.}\ }\textbf
  {\bibinfo {volume} {261}},\ \bibinfo {pages} {451--516} (\bibinfo {year}
  {2006})},\ \Eprint {http://arxiv.org/abs/hep-th/0312085}
  {arXiv:hep-th/0312085 [hep-th]} \BibitemShut {NoStop}%
%%CITATION = HEP-TH/0312085;%%
\bibitem [{\citenamefont {Altshuler}\ \emph {et~al.}(1997)\citenamefont
  {Altshuler}, \citenamefont {Gefen}, \citenamefont {Kamenev},\ and\
  \citenamefont {Levitov}}]{altshuler1997quasiparticle}%
  \BibitemOpen
  \bibfield  {author} {\bibinfo {author} {\bibfnamefont {Boris~L}\ \bibnamefont
  {Altshuler}}, \bibinfo {author} {\bibfnamefont {Yuval}\ \bibnamefont
  {Gefen}}, \bibinfo {author} {\bibfnamefont {Alex}\ \bibnamefont {Kamenev}}, \
  and\ \bibinfo {author} {\bibfnamefont {Leonid~S}\ \bibnamefont {Levitov}},\
  }\bibfield  {title} {\enquote {\bibinfo {title} {Quasiparticle lifetime in a
  finite system: A nonperturbative approach},}\ }\href@noop {} {\bibfield
  {journal} {\bibinfo  {journal} {Physical review letters}\ }\textbf {\bibinfo
  {volume} {78}},\ \bibinfo {pages} {2803} (\bibinfo {year}
  {1997})}\BibitemShut {NoStop}%
\bibitem [{\citenamefont {Basko}\ \emph {et~al.}(2006)\citenamefont {Basko},
  \citenamefont {Aleiner},\ and\ \citenamefont {Altshuler}}]{basko2006metal}%
  \BibitemOpen
  \bibfield  {author} {\bibinfo {author} {\bibfnamefont {Denis~M}\ \bibnamefont
  {Basko}}, \bibinfo {author} {\bibfnamefont {Igor~L}\ \bibnamefont {Aleiner}},
  \ and\ \bibinfo {author} {\bibfnamefont {Boris~L}\ \bibnamefont
  {Altshuler}},\ }\bibfield  {title} {\enquote {\bibinfo {title}
  {Metal--insulator transition in a weakly interacting many-electron system
  with localized single-particle states},}\ }\href@noop {} {\bibfield
  {journal} {\bibinfo  {journal} {Annals of physics}\ }\textbf {\bibinfo
  {volume} {321}},\ \bibinfo {pages} {1126--1205} (\bibinfo {year}
  {2006})}\BibitemShut {NoStop}%
\bibitem [{\citenamefont {Moser}(1975)}]{moser1975finitely}%
  \BibitemOpen
  \bibfield  {author} {\bibinfo {author} {\bibfnamefont {J{\"u}rgen}\
  \bibnamefont {Moser}},\ }\bibfield  {title} {\enquote {\bibinfo {title}
  {Finitely many mass points on the line under the influence of an exponential
  potential--an integrable system},}\ }in\ \href@noop {} {\emph {\bibinfo
  {booktitle} {Dynamical systems, theory and applications}}}\ (\bibinfo
  {publisher} {Springer},\ \bibinfo {year} {1975})\ pp.\ \bibinfo {pages}
  {467--497}\BibitemShut {NoStop}%
\end{thebibliography}%

\clearpage

\section{Supplemental Materials}

\subsection{Toda miscellanea}
Here we mention certain standard results about Toda chain, which are used in the other parts of the paper. 
\subsubsection{Toda EOM in Lax form}
In \eqref{Q} we introduce an orthogonal matrix $Q$ which maps between the family of orthonormal bases $A^t_{n}(t/2)e^{-q_n(t)/2}$, parametrized by $t$, all being associated with the same scalar product $\langle\, \, ,\, \rangle$. Matrix $M(t)$ is simply the matrix of the adjoint action $[H,\, \, ]$ written in the $t$-basis. From here follows that the $t$-dependence of $M(t)$ is an isospectral deformation, 
\bea
M(t)=Q^T(t/2) M(0) Q(t/2). 
\eea
Written in the differential form this becomes Toda equation of motion written in the Lax form  \cite{symes1980hamiltonian},
\bea
\label{QQ}
&&\dot{M}(t)=[B(t),M(t)],\quad \dot{Q}^T(t)=2B(2t) Q^T(t),\\
&&B={1\over 2}\left(\begin{array}{cccc}
0 & b_0 & 0 & \ddots  \\
-b_0 & 0 & b_1 & \ddots   \\
0 & -b_1 & 0 &  \ddots  \\
\ddots  & \ddots &  \ddots & \ddots
\end{array}\right).
\eea

\subsubsection{Hankel determinant representation}
Tau-functions of Toda hierarchy $\tau_n={\rm exp}(\sum_{0\leq k\leq n} q_n)$ can be expressed  concisely in terms of $C(t)=e^{q_0(t)}$ and its derivatives. Namely we introduce $(n+1)\times (n+1)$, $n\geq 0$, matrix 
\bea
{\mathcal M}^{(n)}_{ij}=C^{(i+j)}(t),
\eea
where $C^{(k)}(t)$ stands for $k$-th derivative of $C$. Then 
\bea
\label{Hankel}
\tau_n={\rm det}\, {\mathcal M}^{(n)}.
\eea

\subsection{Continued fraction representation}
Continued fraction representation of the Green's function, 
\bea
{\rm G}(z)=\int_0^{\infty} e^{-z t}\, C(t)\, dt,
\eea
is the central part of the recursion method. From the definition above and representation \eqref{originalGt} we readily find
\bea
\label{Gz}
{\rm G}(z)=\left({1\over z\, {\mathcal I}-M}\right)_{00},
\eea
provided the original operator is normalized, $C(0)=0$.
When matrix $M$ is infinite, the inverse matrix $(z\, {\mathcal I}-M)^{-1}$ should be understood in the formal sense. It is convenient to consider $M$ to be finite, such that $a_n$ are defined for $0\leq n \leq N+1$ and $b_n$ for $0\leq n\leq N$. Then we introduce $M^{(n)}$ as the  $(N-n)\times (N-n)$  bottom-right corner submatrix of $M$. By $\Delta_n$ we denote characteristic polynomial of $M^{(n)}$,
\bea
\Delta_n={\rm det}(z\, {\mathcal I} - M^{(n)}).
\eea
Then ${\rm G}(z)=\Delta_1/\Delta_0$.

To obtain the continued fraction representation we notice that $\Delta_n$ satisfy the following iterative relation
\bea
\Delta_{n}=(z-a_n)\Delta_{n+1}-b_n^2\, \Delta_{n+2}.
\eea
If one defines $s_n=\Delta_{n}/\Delta_{n+1}$, then 
\bea
s_n=(z-a_n)- b_n^2/s_{n+1}.
\eea
From here follows
\bea
\nonumber
{\rm G}(z)\equiv {1\over s_0}= {1\over z-a_0-{b_0^2\over s_1}}={1\over z-a_0-{b_0^2\over z-a_1-{b_1^2\over s_2}}}=\dots
\eea

Continued fraction representation plays an important role in the context of Toda chain as well. In this case
${\rm G}(z,t)$ is defined by \eqref{Gz} with the $t$-dependent $M(t)$. Using \eqref{main} we readily find
\bea
{C(t+s)\over C(t)}=\left(e^{s M(t)}\right)_{00},
\eea
and therefore 
\bea
{\rm G}(z,t)={\int_t^\infty C(t') e^{-z t'} dt'\over C(t)}.
\eea
Green's function ${\rm G}(z,t)$ can be written in terms of the eigenvalues $\lambda_i$ of $M$  and non-negative $r_n$, $\sum_n r_n^2=1$,
\bea
{\rm G}(z,t)={\sum_n  {r_n^2\over z-\lambda_n} \over \sum_n r_n^2}.
\eea
Then time dependence of ${\rm G}$ is described by the gradient flow \cite{moser1975finitely}
\bea
{d\lambda_k\over dt}=0,\quad {d r_k\over dt}=-{\partial V \over \partial r_k},\
V={\sum_n \lambda_n r_n^2\over 2\sum_n r_n^2}.
\eea

\subsection{Exact solutions}
In this subsection we find several families of exact solutions of the Toda chain which exhibit different characteristic behavior: ``chaotic'' $a_n,b_n\propto n$ and ``integrable'' $a_n,b_n \propto n^{1/2}$.
First, we notice that the ``center of mass'' coordinate $\sum_n q_n$ and total momentum $\sum_n \dot{q}_n$ of the Toda chain are free parameters. Hence a transformation $q_n(t)\rightarrow q_n(t)+ v t+{\rm q}$ turns a  solution into a solution, while transforming 
\bea
a_n(t)\rightarrow a_n(t)+v,\quad b_n(t)\rightarrow b_n(t). \label{symm1}
\eea 
Since the Toda equations are not explicitly time-dependent, if $q_n(t)$ is a solution, then $q_n(t-t_0)$ for arbitrary $t_0$ is a also a solution. Finally, rescaling $t$ yields
\bea
q_n(t)&\rightarrow& q_n(Jt)+2k\ln(J),\\
a_n(t)&\rightarrow &J a_n(Jt),\quad b_n(t)\rightarrow J b_n(Jt).
\eea
\subsubsection{``Chaotic'' solutions}
Keeping these symmetries in mind we proceed to construct the family of exact solutions as follows: we use the anzats $b^2_n=b^2(t) p(n)$ where $p(n)$ is a quadratic polynomial. The constant term in $p(n)$ is arbitrary due to \eqref{symm1}. The overall coefficient can be reabsorbed into $b^2$, wile the constant term is fixed by consistency. The most general solution within this anszats is $p(n)=(n+c)(n+1)$ with some $c$. Plugging this into \eqref{eomb} we find
\bea
{d^2\over dt^2} \ln(b^2)=2b^2,\quad b^2={J^2\over \sin^2(J(t_0-t))}.
\eea
This leads to the solution \eqref{exacts},
\bea
\nonumber
\tau_n&=&{G(n+2)G(n+1+c)\over G(c)\Gamma(c)^{n+1}} {J^{n(n+1)}\over \sin(J (t_0-t))^{(n+c)(n+1)}},\\
\nonumber
q_n(t)&=&2n\ln(J)-(2n+c)\ln(J\sin(J(t_0-t)))+ \\ 
\label{exs}
\qquad && \ln(n! \Gamma(n+c)),\\
\nonumber
a_n(t)&=&(2n+c)J \cot (J(t_0-t)),\\
b^2_n(t)&=&{ (n+c)(n+1) J^2 \over \sin^{2}(J(t_0-t))}, \nonumber
\eea 
where $G(x)$ is the Barnes  gamma function. Positivity of $b_0^2(t)$ requires $c\geq 0$.

After taking the  limit $J\rightarrow 0$ and using the symmetry \eqref{symm1} the solution becomes
\bea
\label{Zs}
\tau_n&=&{G(n+2)G(n+1+c)\over G(c)\Gamma(c)^{n+1}} {1\over (t_0-t)^{(n+c)(n+1)}},\\
\label{Zsq}
q_n&=&-(2n+c) \ln(t_0-t)+\ln(n!\, \Gamma(n+c)),\\
\label{Zsab}
a_n&=&{2n+c\over t_0-t},\quad b_n^2={(n+c)(n+1)\over (t_0-t)^2}.
\eea

The family of solutions  \eqref{exs} can be further analyzed. We would like to find the explicit form of the orthogonal transformation $Q(t)$. From \eqref{QQ} it follows that each row of $Q$, which we (somewhat surprisingly) denote by $\psi$, willl satisfy
\bea
\label{psi}
\dot{\psi}(t)=2B(2t)\psi(t).
\eea
Using explicit form of $b_n(t)$ we factor out time-dependence of $B(t)$,
\bea
2B(2t)={1\over \sin(J(t_0-2t))} 2B(t_0).
\eea
It is convenient to  introduce auxilary ``time'' variable $t_M(t)$ which satisfies $d t_M/dt=J/\cos(J(t_0-2t))$,
\bea
\label{Minkowski}
J t_M={1\over 2}\ln {\cot(J(t_0/2-t))\over \cot(J(t_0/2))}.
\eea 
Then $\psi(t_M(t))$ will solve \eqref{psi}, provided ${d\psi/dt_M}=2B(t_0) \psi(t_M)$. 
Since $a_n(t_0)=0$, matrix $2B(t_0)$ is related by a simple unitary transformation to $i M(t_0)$. Therefore, up to a trivial facor, $\psi(t_M)$ describes conventional (Minkowski) time evolution of an operator in Krylov space.
This explains the choice of notations for $\psi$ -- the ``wave-function'' of the operator, and $t_M$ -- time in Minkowski space. 
For the system described by Lanczos coefficients $a_n=0, b_n=(n+c)(n+1)$, a particular solution with $\psi_n(0)=\delta_{n0}$  was found in \cite{Parker_2019},
\bea
\label{psisol}
\psi_n(t_M)=(-1)^n\sqrt{\Gamma(n+c)\over n!\, \Gamma(c)} {\tanh^n(J\, t_M)\over \cosh^c(J\, t_M)}.
\eea
Since $2B(t_0)$ is time-independent, other solutions can be  obtained by acting  on  \eqref{psisol}  by  differential operators with constant coefficients, e.g. $\psi^{(1)}=c^{-1/2}\psi'(t_M)$ is a solution satisfying $\psi^{(1)}_n(0)=\delta_{n1}$. After substituting  \eqref{Minkowski} as an argument of $\psi$, it becomes first  row of matrix $Q$, 
\bea
\nonumber
Q_{n0}(t)=(-1)^n \sqrt{\Gamma(n+c)\over n!\, \Gamma(c)}\left({\sin(J t_0)\sin(J (t_0-2t))\over \sin^2(J (t_0-t))}\right)^{c/2} \times \\ \times \left({\sin(J t)\over \sin(J(t_0-t))}\right)^n,\qquad \quad 
\label{Qex}
\eea
while $\psi^{(1)}$ will become second row, etc.

From the explicit solution it is easy to see that at $t=t_0/2$ all components of $Q_{n0}$ vanish, while the product $R_{00} Q_{n0}$ is regular. In fact all components $Q_{nm}$ vanish at $t=t_0/2$. This is easy to see by going back to the ``Minkowski'' time $t_M$ \eqref{Minkowski}. When $t\rightarrow t_0/2$, $t_M\rightarrow \infty$. In this limit all components of \eqref{psisol} decay exponentially. Since all rows of $Q$ can be obtained by acting on $\psi_n(t_M)$ by a differential operator with constant coefficients, they all will decay exponentially with $t_M$ and therefore vanish at $t=t_0/2$.

Using explicit solution \eqref{Qex} one can easily calculate the  inverse participating ratio \eqref{IPR} to immediately conclude that it diverges at $t=t_0/2$. 

The behavior of  \eqref{Qex}  near $t=t_0/2$ is typical  where $t_0=t^*$ is the point of singularity. To show that we assume that near $t=t^*$ the tau-function behaves as 
\bea
\tau_0 \propto {1\over (t^*-t)^c}.
\eea 
Using Hankel determinant representation \eqref{Hankel} we immediately find that the singular behavior of $\tau_n$ near $t=t^*$ is given by \eqref{Zs} with $t_0=t^*$, from where the singular behavior of $q_n,a_n,b_n$  near $t=t^*$ given by (\ref{Zsq},\ref{Zsab}) follows.

From the identity $R_{00}(t/2)^2=\tau_0(t)/\tau_0(0)$ one immediately sees that near $t=t^*/2$, $R_{00}(t)\propto (t^*-2t)^{-c/2}$, and from $R_{00}(t)Q_{00}(t)=\tau_0(t)/\tau_0(0)$ and regularity of $\tau_0(t)$ at $t=t^*/2$ one concludes 
\bea
Q_{00}(t) \propto (t^*-2t)^{c/2},
\eea
near $t=t^*/2$. Now one can use the differential equation for $Q$ \eqref{QQ}, $\dot{Q}_{00}(t)=b_0(2t) Q_{10}(t)$, together with the leading singular behavior of $b_0$ near $t=t^*/2$  \eqref{Zsab} to conclude that  $Q_{10}(t) \propto (t^*-2t)^{c/2}$, and so on.

\subsubsection{``Integrable'' solutions}
There is an exact family of solutions $b_n^2=b^2(t) p(n)$, where $p(n)$ is a linear function of $n$. Without loss of generality we can choose $p(n)=n+c$, and later see that self-consistency requires $c=1$. Then $b^2=e^{m(t-t_0)}$, and 
\bea
\nonumber
\tau_n&=& G(n+2)e^{{(n+1)\over m^2} e^{m(t-t_0)}}e^{m(n+1)(n+2)(t-t_0)/2},\\ \nonumber
q_n&=&{e^{m(t-t_0)}\over m^2}+(n+1)m(t-t_0)+\ln (n!),\\
a_n&=&e^{m(t-t_0)}, \qquad b^2_n=(n+1)e^{m (t-t_0)}. \label{1d}
\eea 
From the positive of $b_0^2$ follows $c\geq 0$.

In the limit $m\rightarrow 0$ exponent $e^{m(t-t_0)}$ can be expanded in Taylor sereis and after rescaling of $t$ one finds, 
\bea
\tau_n&=& G(n+2) e^{n(n+1)/2 \ln(a)} e^{a(n+1)(t-t_0)^2/2},\\
q_n &=& {a(t-t_0)^2\over 2}+n\ln(a)+\ln(n!),\\
a_n &=& a (t-t_0),\qquad b_n^2=a (n+1).
\eea
Since $b_n^2$ are time independent the differential equation for $Q_{n0}$ is particularly easy to solve 
\bea
Q_{n0}={(-a^{1/2}t)^n\over \sqrt{n!}}e^{-at^2/2}.
\label{Gauss}
\eea
This is a ``wave-packet'' centered at $n \sim t^2$.
It is also easy to calculate  inverse participation ratio \eqref{IPR},
$I={e^{2 a t^2}/ I_0(2 a t^2)}$, which grows linearly with $t$.

Going back to the solution \eqref{1d}, the ``wave-function'' $Q_{n0}$ is given by \eqref{Gauss} with $t$ substituted by 
\bea
a^{1/2}t\rightarrow {e^{-m t_0}\over m}\left(e^{m t}-1\right),
\eea
which means inverse participation ration grows exponentially with $t$.
\end{document}